**Long-range Ni/Mn structural order in epitaxial double perovskite $La_2NiMnO_6$ thin films**

M.P. Singh*, K.D. Truong, S. Jandl, and P. Fournier

Regroupement québécois sur les matériaux de pointe, Département de physique, Université de Sherbrooke, Sherbrooke, J1K 2R1 Canada

We report and compare the structural, magnetic, and optical properties of ordered $La_2NiMnO_6$ thin films and its disordered $LaNi_{0.5}Mn_{0.5}O_3$ counterpart. An x-ray diffraction study reveals that the B-site Ni/Mn ordering induces additional XRD reflections as the crystal symmetry is transformed from a pseudocubic perovskite unit cell in the disordered phase to a monoclinic form with larger lattice parameters for the ordered phase. Polarized Raman spectroscopy studies reveal that the ordered samples are characterized by additional phonon excitations that are absent in the disordered phase. The appearance of these additional phonon excitations is interpreted as the clearest signature of Brillouin zone folding as a result of the long-range Ni/Mn ordering in $La_2NiMnO_6$. Both ordered and disordered materials display a single ferromagnetic-to-paramagnetic transition. The ordered films display also a saturation magnetization close to 4.8 $\mu_B$/f.u. and a transition temperature (FM-$T_C$) around 270 K, while the disordered ones have only a 3.7 $\mu_B$/f.u. saturation magnetization and a FM-$T_C$ around 138 K. The differences in their magnetic behaviours are understood based on the distinct local electronic configurations of their Ni/Mn cations.

**\* Corresponding author:** *mangala.singh@usherbrooke.ca*

## Introduction

The coexistence of ferromagnetic, insulator and polar behaviours in double perovskite $La_2B'B''O_6$ (say $B'$ = Mn; $B''$ = Co, Ni) is of great interest in designing innovative multiferroics and to explore the underlying multiferroic coupling behavior in oxides[1-24]. The demonstration of a large magnetodielectric effect in bulk $La_2NiMnO_6$ (LNMO)[3] and $La_2CoMnO_6$ (LCMO) thin films[4] in the vicinity of their respective magnetic transition close to room temperature reveals the far-reaching technological potential of double perovskites. However, this large effect observed close to their transition temperature (FM-$T_c$) seems to contradict the magnetodielectric effect of $La_2NiMnO_6$ thin films[5] observed far below their magnetic transition temperature. One of the possible reasons for this different behaviour may be related to the absence of long-range structural order. In the abovementioned LNMO bulk and LCMO thin films, the Ni (Co) and Mn ions alternate along specific crystallographic directions over very long distances[24]. In the LNMO thin films[19], it was shown that such structural order is limited only to small domain sizes of about 50-100 nm. These studies suggest also that the B-site ordering may play a crucial role in determining the physical properties of double perovskites, in particular the coincidence of the large magnetoelectric response and the magnetic transition.

In fully-disordered LNMO, with chemical formula $LaNi_{0.5}Mn_{0.5}O_3$, the Ni and Mn cations are randomly distributed at the 3d-metal (B) sites of a typical pseudocubic $ABO_3$ perovskite unit cell. This random distribution results in Ni-O-Mn, Ni-O-Ni, and Mn-O-Mn chemical bonds with different superexchange interactions. Ni-O-Ni and Mn-O-Mn bonds are expected to give paramagnetic and antiferromagnetic superexchange interactions respectively[6-8]. Their presence hinders the formation of high-temperature long-range ferromagnetic order arising from the ferromagnetic interaction of the Ni-O-Mn bonds, leading instead to low FM-$T_c$. In long-range ordered LNMO (with unit-cell formula $La_2NiMnO_6$), only Ni-O-Mn bonds are possible leading to the highest FM-$T_c$. Furthermore, the detailed bulk studies using various techniques[3, 6-14] (e.g., $^{55}$Mn NMR, X-ray photoelectron spectroscopy, X-ray absorption spectroscopy, and neutron scattering) have clearly shown that the ordered phase includes only $Ni^{2+}$ and $Mn^{4+}$ oxidation states, while the disordered phase has $Ni^{3+}$ and $Mn^{3+}$ ions. [6-20] The ordering of $Ni^{2+}$ and $Mn^{4+}$ at the B-sites is also expected to induce a local polar behaviour similar to LCMO and the charge ordered manganites [4]. Thus, such charge ordered system should present an important impact on phonon dispersion and magnetic transition, and eventually coupling to the magnetic order similar to LCMO and other double perovskites [17-18, 25-27].

It is therefore interesting to compare the physical properties of the ordered and disordered LNMO systems and establish a clear relationship between the structural and other functional properties. Unlike the bulk, only very few efforts have recently been made to study the structural and functional properties of LNMO thin films [15-20]. Singh *et al.* [19] have shown that the films displaying multiple magnetic transitions do possess both ordered and disordered phases at room temperature. The appearance of a second magnetic phase transition below 200 K was related to the disordered parts of the films [19]. On the other hand, Gupta and coworkers have recently shown that the optimization of the growth parameters can lead to a single magnetic transition [16] as confirmed also recently by Kitamura et al.[20] However, subsequent Raman studies and transmission electron microscopy studies demonstrated that these films present only short-range Ni/Mn ordering as the expected additional phonons of the long-range ordered phase could not be seen[16-18] contrary to LCMO [22]. Moreover, these films display the strong spin-lattice coupling and magnetodielectric effect far below their magnetic Curie temperature. [3-5, 17]

The abovementioned studies on thin films clearly illustrate that these LNMO thin film samples have either shown the presence of a short-range ordered phase in LNMO or the coexistence of multiple phases similar to the polycrystalline bulk samples [15-20]. Actually, the long-range ordered $La_2NiMnO_6$ and fully disordered $LaNi_{0.5}Mn_{0.5}O_3$ phases have yet to be stabilized separately in thin film forms. Such situation also hinders our ability to understand their respective functional properties. In this paper, we compare the physical properties of long-range ordered and fully disordered LNMO thin films made by pulsed-laser deposition. We present the growth parameters that promote the ordering in LNMO and obtain the stability phase diagram underlining a very narrow growth parameter window compared to LCMO. Our experimental results illustrate that the ordered and disordered LNMO films have distinct electronic and magnetic properties in particular different magnetic transition temperatures, low temperature saturation magnetizations and phonon dispersion relations.

## Experimental details

The LNMO epitaxial films were grown on (111)- and (001)-SrTiO$_3$ [respectively STO (111) and STO (001)] using the pulsed-laser deposition (PLD) technique. While previous studies have been reported mainly on the growth of these films on STO (001) [15-20], it is interesting to focus instead on the growth of LNMO on STO (111) since one expects ordering of the NiO and MnO planes along the out-of-plane growth direction. The epitaxial growth stability and quality was explored in the temperature range of 475-850 °C and under an oxygen pressure range of 100-900 mTorr. Following the deposition, the growth chamber was filled with 400 Torr of O$_2$ and the films were cooled down to room temperature at a rate of 10 °C/min. For ablation, a polycrystalline stoichiometric LNMO target was synthesized using the standard solid-state chemistry route. Crystallinity and epitaxial quality of our films were studied using X-ray diffraction (XRD) in the θ-2θ and rocking (ω) curve modes. The Raman scattering measurements were carried out at 10 and 295 K in the XX, XY, and X'Y' and X'X' scattering configurations for films grown on SrTiO$_3$ (111) using a polarized Raman spectrometer in the backscattering mode with a CCD camera giving a 0.5 cm$^{-1}$ resolution [26]. Finally, the temperature and field dependence of the magnetization were measured using a Superconducting Quantum interferometer devices (SQUID) magnetometer from Quantum Design.

## Results and Discussion

The ideal ordered La$_2$NiMnO$_6$ double perovskite system[24] can be described in terms of an ideal ABO$_3$ perovskite unit cell wherein Ni/Mn cations at the B-sites are ordered along the (111)$_P$ planes (where "P" stands for pseudocubic unit cell). In a long-range ordered system, Ni and Mn sublattices are thus occupied only by Ni or Mn cations, respectively. On the contrary, Ni and Mn are randomly distributed in a fully disordered system. A short-range ordered sample would be characterized by a partial ordering of Ni and Mn sublattices at the B-sites over a limited domain size. The important feature that distinguishes essentially the Ni and Mn ordering in bulk La$_2$NiMnO$_6$ is a change of crystal symmetry as was previously reported [3, 7, 12, 13]. The ordered bulk LNMO displays a monoclinic P2$_1$/n symmetry with a unit cell parameters a = 5.515 Å (a$_{P1}$ = a/√2 ~ 3.911Å), b = 7.742 Å (a$_{P2}$ = b/2 ~ 3.871Å), c = 5.46 Å (a$_{P3}$ = c/√2 ~ 3.872Å), and β = 90.04°. On the other hand, the disordered phase possesses an orthorhombic Pbnm symmetry with lattice parameters a = 5.50 (a$_P$ = a/√2 ~ 3.90Å) Å, b = 5.54 Å (a$_P$ = b/√2 ~ 3.929Å), c = 7.735 Å (a$_P$ = a/2 ~ 3.868Å) at room temperature. a$_P$ stands for the lattice parameter of an ideal pseudocubic unit cell.

Figure 1 focuses on the θ-2θ patterns of films grown on STO (111) under conditions leading to fully disordered LaNi$_{0.5}$Mn$_{0.5}$O$_3$ and long-range ordered La$_2$NiMnO$_6$ films in (a) and (b) respectively. The cation-disordered film in Fig. 1a is grown at 500 °C and an O$_2$ pressure of 300 mTorr while the ordered film in Fig. 1b is grown at 800 °C and an O$_2$ pressure of 800 mTorr. Irrespective of the growth conditions, the films are always displaying a diffraction peak in very close proximity to the STO (111) reflection attesting that the films are growing coherently with 3d-metal interatomic distances matching closely those of STO. To verify further the coherence of our films, we performed the rocking curves. The well-ordered film is found to be characterized by 0.35° full-width at half maximum (FWHM) confirming the coherent nature of our films. There are two ways we can index the observed XRD reflections. First in the orthorhombic or monoclinic crystal symmetries [3, 12, 24], these peaks can be indexed as the (022) or (220) reflections leading to an average lattice parameter of a$_P$ ~ 3.86 Å. This indicates that the LaMnO$_3$ and LaNiO$_3$ primitive cells still lie with their (111) directions along that of STO (111).

More importantly, the films grown at 800 °C / 800 mTorr display additional XRD peaks (Fig. 1b). Three peaks may be indexed as (101)$_m$, (002)$_m$ and (202)$_m$ and can barely be seen in films grown at low temperature. The subscript "m" represents the monoclinic symmetry of LNMO. This demonstrates that the films grown at the highest temperatures possess a monoclinic symmetry and that several domain orientations may be stable on STO (111) by a subtle relative change in the lattice distortions expected from the monoclinic symmetry. Furthermore, it illustrates a clear B-site ordering that provokes a transition from the high symmetry averaged cubic unit cell of the disordered phase to the monoclinic structure of the monoclinic phase [24]. In such case, the LNMO films grow with alternating $NiO_2$ and $MnO_2$ planes along the (111) direction of STO with an interplanar distance of roughly 2.2 Å corresponding to half of the chemical modulation wavelength (of 4.4 Å). Our films on STO (111) grown at high temperatures also exhibit one additional reflection, marked as *. We have looked into various possibilities to index this peak including the possible presence of an impurity phase such as $LaMnO_3$, $LaNiO_3$ or NiO. The lattice parameters and crystal symmetry of these phases match closely with LNMO. However, the origin of this reflection arising from an impurity phase can be easily ruled out based on the magnetic and Raman properties of these films.

A second appealing scenario based on the presence of superlattice reflections may explain the extra peaks in the x-ray diffraction pattern of the ordered LNMO. It is important to note that as we approach close to the main (022) reflection, the intensity of the *-denoted, (002) and (202) reflections grows progressively. It thus presents the typical characteristics of a superlattice structure [24]. From this perspective, the (202) reflection may be also indexed as a +1 satellite peak, the (002) reflection as a -1 satellite peak and the * reflection as a -2 satellite peak arising from additional Ni/Mn ordering [24]. Assuming such scenario, we estimated the corresponding chemical modulations ($\Lambda$) of our films using $\Lambda = n\lambda/2[\sin\theta_i - \sin\theta_{i-1}]$, where $\lambda$ is the X-ray wavelength, $\theta_i$ is the peak position of $i^{th}$ satellite peak, and $n^{th}$-order. The estimated average value of $\Lambda'$ (between 0 and +1 and as well as 0 and -2 satellite peak) and $\Lambda''$ (0 and -1 reflection) are about 14.75 Å and about 30 Å, respectively. The $\Lambda'$ is very close to the theoretical value of 14 Å for an ideal $ABO_3/AB'O_3$ superlattice. The modulation of 30 Å may arise from an additional out-of-phase rotation of the Ni/Mn octahedra from one unit cell to the other in a consecutive manner. This would explain the origin of the *-denoted reflection. Such additional rotation may be necessary to minimize further the elastic energy arising from the local polarization caused by the long-range ordering and the monoclinic distortion. Tilting of the octahedra is also observed in other well-ordered double perovskites [24] (*e.g.,*

La$_2$CoMnO$_6$ [22] and Ca$_2$FeReO$_6$ [23]) leading to a unique structural domain growth. [22-24] Additional structural studies, such as transmission electron microscopy and synchrotron x-ray scattering are clearly warranted to probe and confirm these octahedral distortions.

Cation ordering is found to be very sensitive to the temperature and the oxygen pressure during the growth process. As an illustration, we present in the inset of Fig. 1a the value of the b-axis lattice parameter extracted from the *(0l0)* XRD peaks of the films grown on STO (100) at 800°C as a function of oxygen growth pressure. The expansion of the lattice is a direct consequence of the extension of the b-axis as long-range structural order sets in. Because of the presence of a large local polarization field created due to alternating planes of Ni$^{2+}$ and Mn$^{4+}$ ions along the (101) direction (and equivalent directions) in the long-range ordered phase, the LNMO lattice will search to minimize the extra coulombic energy by maximizing the unit cell volume. This observed expansion of the b-axis lattice parameter with cation-ordering is consistent with the other double perovskite films [22].

Relying on these structural and physical properties as signatures of the level of ordering in our films, we can map out the phase-stability diagram (Fig. 2) for the long-range ordered LNMO (shaded elliptical area) and disordered LNMO (shaded rectangular area). We compare also our growth conditions and film characteristics with the available literature (Fig. 2) [16-19]. We find that the epitaxial films grown in the vicinity of 500 °C present only the fully disordered phase with no trace of impurity phases. Those grown in a narrow temperature window of only ±20 °C around 810 °C and about 800 mTorr have all the expected physical properties of the ordered phase. A deviation from these growth parameters leads to an admixture of both phases (*i.e.,* ordered and disordered) in different proportions. The observed phase-stability zone for the ordered LNMO is consistent with the phase-stability diagrams of other double perovskite films, such as La$_2$CoMnO$_6$ and Sr$_2$FeMoO$_6$. [21, 22] Interestingly, we can note that the stability window for ordered LNMO is much narrower than for LCMO [22]. Thus, our study shows that the ordered B-site configuration in double perovskite systems can be stabilized only by growing the films at relatively high temperature and under relatively high oxygen pressure, but that the exact size of this zone in the phase diagram may vary for different sets of B ions. However, the same trend observed in the growth of long-range LNMO and LCMO [2, 22] may be considered as a generic signature on the path to stabilise the long-range ordered cationic structure in La-based double perovskites films by PLD.

Polarized Raman spectroscopy is a proven technique to identify local ordering, structural distortions, and spin-lattice coupling in double perovskite systems [25-27]. For example, it has been used recently to demonstrate the presence of long-range B-site order in LCMO films [26], short-range order in LNMO films [17] and a cation-disordered structure for $LaBiMn_{4/3}Co_{2/3}O_6$ films [27]. To determine and compare the local ordering and their likely influence on the phonon excitations in our samples, we measured the temperature and polarization dependences of the Raman spectra on our films. A typical polarization dependence of the Raman spectra measured in the XX, X'X', XY, X'Y' configurations on a well-ordered film are presented in Fig. 2b. The intensities of the phonon excitations show a strong dependence on the polarization configuration that matches closely with theoretical predictions for the monoclinic LNMO and LCMO by Iliev *et al.* (Refs 17, 25). It thus confirms that these films have a monoclinic symmetry.

Raman scattering is the most stringent test to demonstrate long-range cation ordering in double perovskites [17, 25, 26]. To illustrate it, we measured the temperature dependence of the Raman spectra (Fig. 3) in the XY configuration on both films. Strong modes around 650 and 527 cm$^{-1}$ are assigned to stretching and anti-stretching vibrations of the (Ni/Mn)O$_6$ octahedra, respectively. By a careful inspection of their temperature dependence, various important features can be noted. *First,* the disordered films (Fig 3a) are characterized by broad peaks comparable to the ordered films (Fig 3b). *Second,* the ordered films are characterized by a large number of Raman phonon excitations compared to disordered films. For example, only two peaks can be resolved for the disordered films (inset of Fig 3a) around 650 cm$^{-1}$ corresponding to Ni/Mn-O stretching modes while four peaks (inset of Fig 3b) are easily identified in the well-ordered films [17, 25-26]. It clearly illustrates that additional phonon excitations emerge in the well-ordered films.

A detailed analysis of the polarization dependence of the Raman spectra at 10 K shows that our LNMO ordered films present a total of 17 Raman active phonons while the disordered films show only 6 Raman active phonons. The appearance of additional Raman active phonons in the long-range ordered films could be interpreted as follows. Raman spectra of a pseudocubic $ABO_3$ perovskite [25] are characterized by a distinct number of Raman-active phonons and their intensity shows a unique dependence on the polarization configurations. However, when we dope the B-site with another cation B' (Fig 4), it induces local structural distortions that will impact the phonon spectrum in many ways. If B/B' are randomly distributed at the B-sites (Fig. 4a) as in $LaNi_{0.5}Mn_{0.5}O_3$ with little changes in the pseudocubic lattice parameters [24], the number of Raman excitations should remain the same as that of $LaMnO_3$ for example. However, the frequency and the width of the Raman phonon peaks may be altered due to the change in the phonon lifetimes and the phonon density of states. In the case where B and B' cations are alternating in an $ABO_3$ unit cell (Fig. 4b) as in $La_2NiMnO_6$, the lattice parameters increase leading to an effective change in crystal structure [24], as also evidenced by our XRD studies (Fig 1). This provokes a Brillouin zone folding leading to a major impact on the number of phonon modes observed. Experimentally, it is observed as a clear splitting of Raman-active modes (Fig 3). In other words, Brillouin zone folding leads to new Γ-point Raman excitations in the phonon spectra, which would be absent otherwise. Furthermore, long-range Ni/Mn ordering may also enhance the phonon lifetime resulting in a better-resolved peaks for the ordered LNMO films. A similar effect has recently been demonstrated with the LCMO double perovskites [25, 26]. In principle, these behaviours are prominent at any temperature (Fig 3). However, it is easily observable at low temperature due to a drastic reduction in the thermal peak broadening. This explains the emergence of additional phonons in the long-range ordered films. In particular, this also explains the presence of four peaks in the ordered films around 650 cm$^{-1}$ compared to two peaks in the disordered films in the same frequency range (Fig. 3). These features are well consistent with the observed and theoretically predicted behaviour of the ordered LCMO [22] and LNMO [17, 18].

As pointed out earlier, cation ordering plays a crucial role in determining the magnetic properties of LNMO. As a final test, we studied the magnetic behaviour of our films. Typical M-T curves recorded on the ordered and disordered samples are shown in the inset of Fig 5. It clearly shows that both types of films are characterised by a single magnetic transition, but they differ significantly in their ferromagnetic-to-paramagnetic Curie temperature (FM-$T_c$) values. The disordered film has a FM-$T_c$ at 138 K while the ordered films exhibit a single ferromagnetic-to-paramagnetic transition around 270 K. These values are consistent with the literature on bulk samples and show that the disordered films present random distribution of $Ni^{3+}/Mn^{3+}$ whereas the ordered samples have a dominant $Ni^{2+}/Mn^{4+}$ configurations [3, 6-16, 18]. Our study also corroborates previous reports that the low temperature FM-$T_c$ is definitely related to the disordered phases of LNMO [19]. A typical M-H loop recorded at 10 K is shown in Fig. 5. It clearly shows that both types of films are characterized by a well-defined hysteresis loop. The coercivity of the disordered films is found to be roughly 1 kOe while it reaches values as low as 0.25 kOe for the ordered films. The saturation magnetization for the ordered films is found to be close to 4.8 $\mu_B$/f.u., a value substantially higher than that of the disordered films (3.7 $\mu_B$/f.u).

The magnetic interactions in LNMO are governed by the 180°-superexchange process [7, 28] arising from the bond between oxygen and the partially filled Ni ($e_g$) and Mn ($e_g$) ions giving rise to a rich set of magnetic orders tuned by Ni-Mn distance and the Ni-O-Mn bond angle. The electronic configurations of Ni and Mn, set by their possible oxidation states $Ni^{2+}$, $Ni^{3+}$, $Mn^{3+}$, $Mn^{4+}$, play a crucial role in determining the superexchange strength [3, 6-16, 28]. Unlike the case of LCMO [22, 28-29], the saturation magnetization does not allow us to distinguish between the long-range ordered $Ni^{2+}$ ($d^8$: $t_{2g}^6 e_g^2$) and $Mn^{4+}$ ($d^3$: $t_{2g}^3 e_g^0$) and the disordered low-spin $Ni^{3+}$ ($d^7$: $t_{2g}^6 e_g^1$) and high spin $Mn^{3+}$ ($d^4$: $t_{2g}^3 e_g^1$) configurations. In both cases, the theoretical values of the saturation magnetization assuming only Ni-O-Mn bonds are therefore expected to be ~5 $\mu_B$/f.u. We did obtain such value of saturation magnetization in the case of our ordered films. In the case of the disordered films, the large proportion of Ni-O-Ni and Mn-O-Mn bonds leading to antiferromagnetic/paramagnetic superexchange interaction is likely responsible in part for their low value of saturation magnetization. Moreover, the superexchange interaction in the $Ni^{3+}$-O-$Mn^{3+}$ bonds is likely further suppressed since both $Ni^{3+}$ and $Mn^{3+}$ are Jahn-Teller cations that will generate local structural distortions which may further impede the strength of the superexchange interaction. These combined effects are likely responsible for the observed reduction in the value of FM-$T_c$ in disordered LNMO.

In summary, we have studied the functional properties of ordered and disordered $La_2NiMnO_6$. We demonstrate that it is possible to obtain a long-range Ni/Mn ordering in $La_2NiMnO_6$ by growing the materials in large temperatures and oxygen pressures. The cation ordering significantly influences the optical properties and lead to additional phonon excitations caused by Brillouin zone folding. Furthermore, the ordered and disordered samples have distinct superexchange strengths and consequently different magnetic behaviours. Our study may open the possibility to promote the cationic ordering and to control the multifunctional behaviours of double perovskites for potential applications.

## Acknowledgments

We thank S. Pelletier and M. Castonguay for their technical assistance. This work was supported by the Canadian Institute for Advanced Research, Canada Foundation for Innovation, Natural Sciences and Engineering Research Council (Canada), Fonds Québécois pour la Recherche sur la Nature et les Technologies (Québec), and the Université de Sherbrooke.

**Figure captions**

**Figure 1: (Color online)** Typical XRD patterns of LNMO films grown on STO (111) at **(a)** 500 °C and 300 mTorr $O_2$ and **(b)** at 800 °C and 800 mTorr. The inset in (a) presents the out-of-plane lattice parameter of the films grown on STO (001) at 800 °C as a function of the growth pressure. The inset in (b) shows a rocking curve recorded on the (202) reflections.

**Figure 2: (Color online)** The phase-stability diagram of ordered $La_2NiMnO_6$ (elliptical shaded area) and disordered $LaNi_{0.5}Mn_{0.5}O_3$ (rectangular shaded area). The symbols position the growth conditions of our LNMO and some found in the literature: solid circles for the present work; solid squares from Ref. 19; and yellow stars from Refs. 16 and 18). It is important to note that both the ordered and disordered phases can be independently stabilized only in a very narrow parameter window. **(b)** Typical polarization configuration dependence of the Raman spectra for ordered LNMO films at 298 K.

**Figure 3: (Color online)** Raman spectra for the (a) disordered and (b) well-ordered films measured in 10 – 300 K temperature range under XY configuration. The corresponding insets show fits (shaded curves) to the 10 K experimental spectra around 650 $cm^{-1}$ confirming the presence of two and four stretching modes in the disordered and ordered films, respectively.

**Figure 4: (Color online)** Schematics of $A_2B'B''O_6$ crystal structures displaying **(a)** a random distribution of B'/B'' in the pseudo-cubic $ABO_3$ with a fully disordered unit cell, **(b)** alternatively arranged B'/B'' in ideal $ABO_3$ unit cell leading to a long-range ordered double perovskite. The long-range cation ordering redefines the lattice parameter and crystal structure as shown in **(c)**. Red spheres : B' cations; yellow spheres : B'' cations, large spheres : A-cations; and blue spheres : oxygen atoms.

**Figure 5: (Color online)** M-H loops for **(a)** a disordered film and **(b)** an ordered film at 10 K. The corresponding insets show the respective M-T curves displaying the distinct single *ferromagnetic-to-paramagnetic* phase transitions.

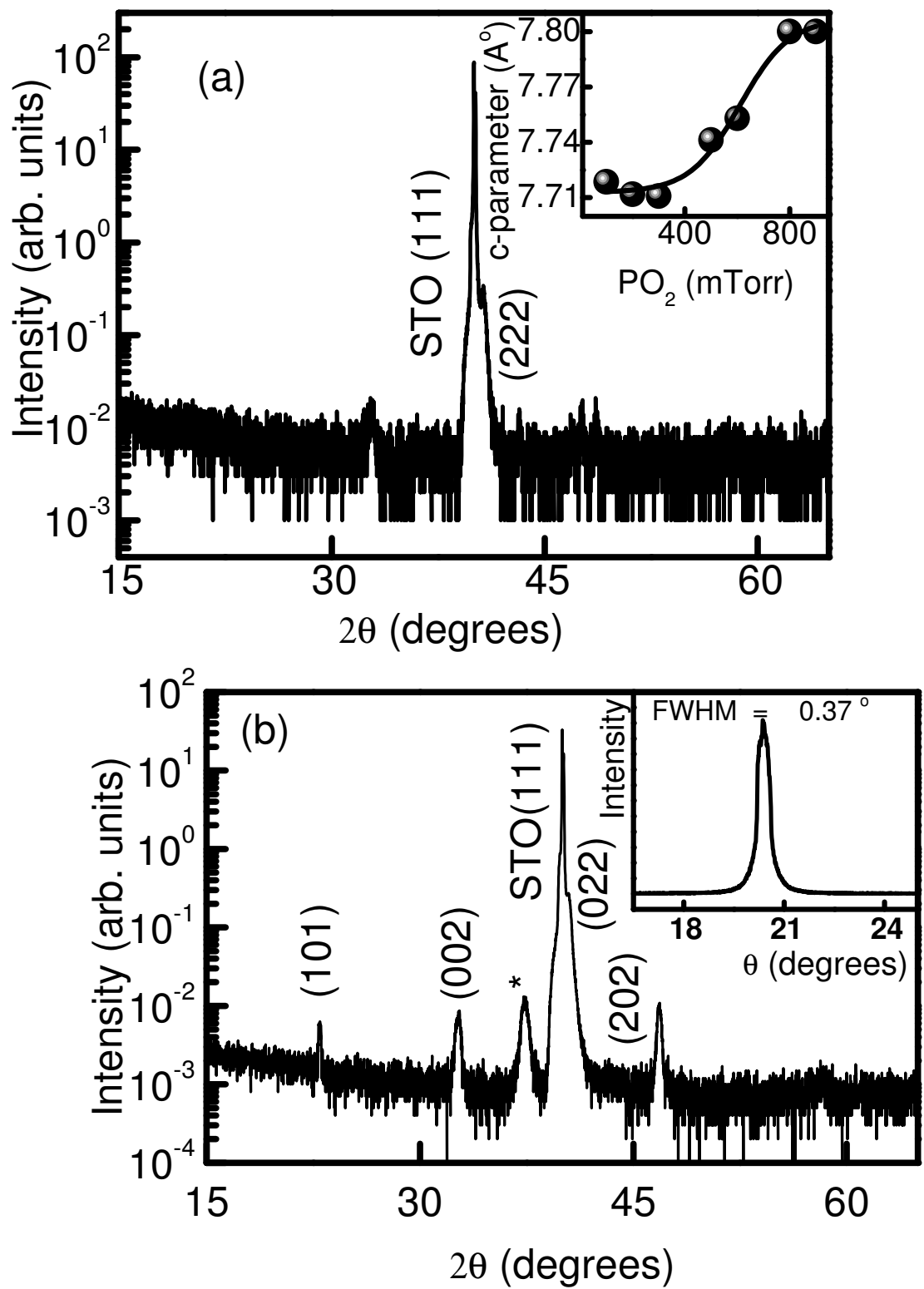

Figure 1

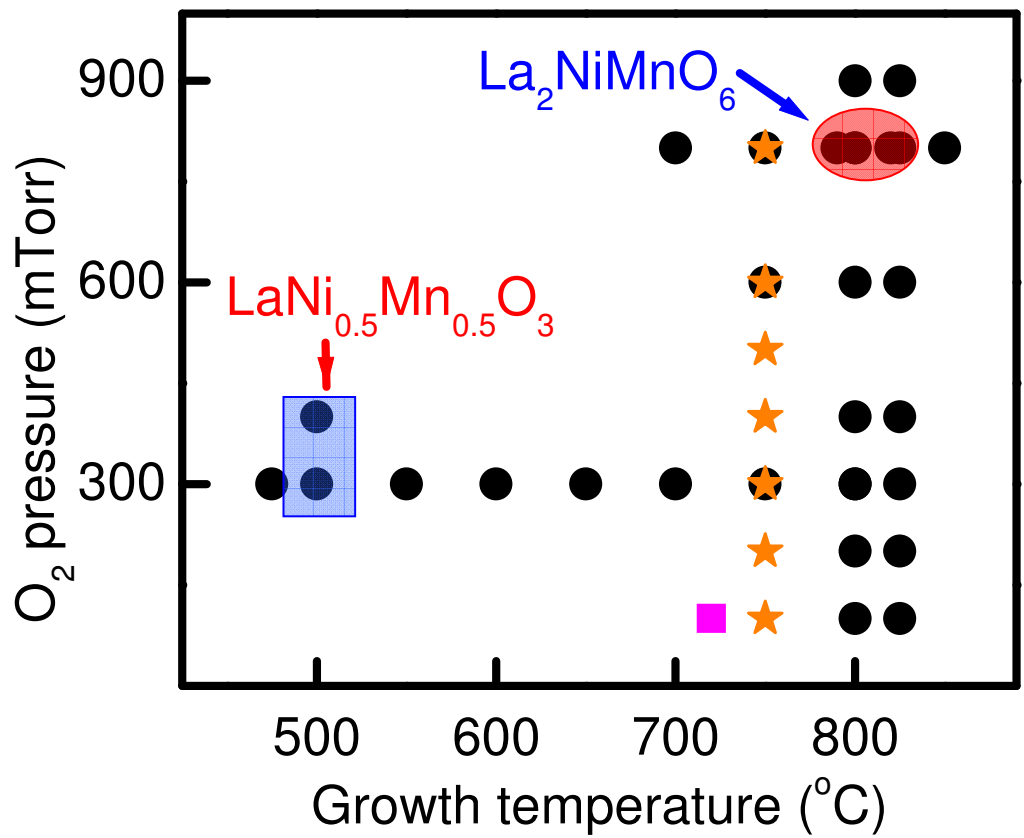

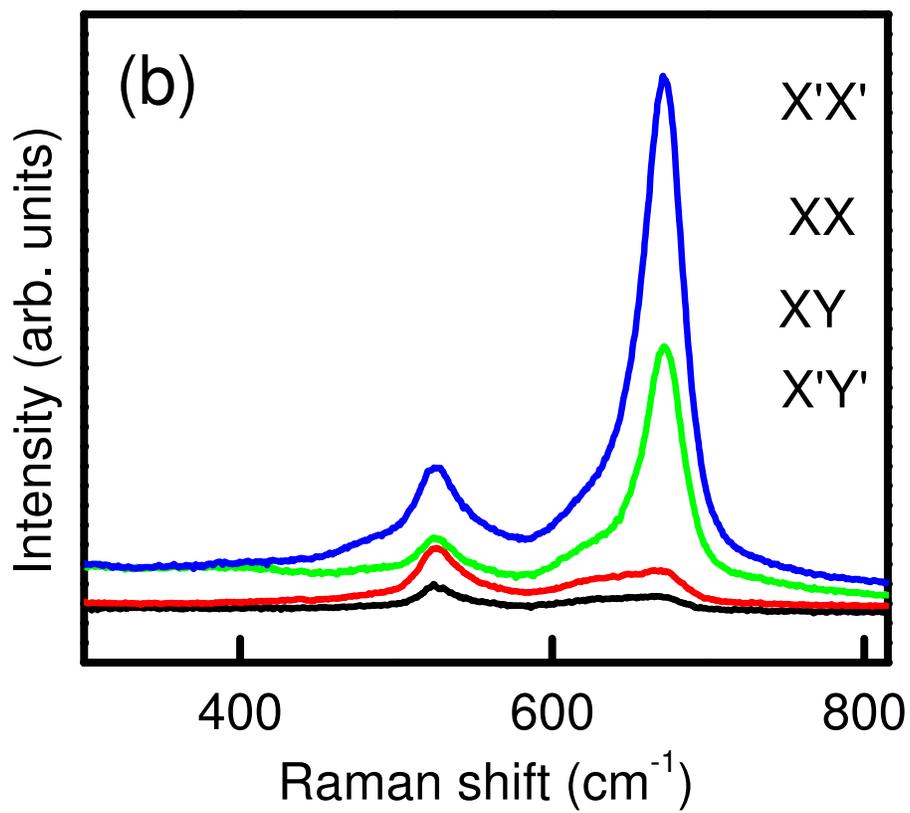

Figure 2

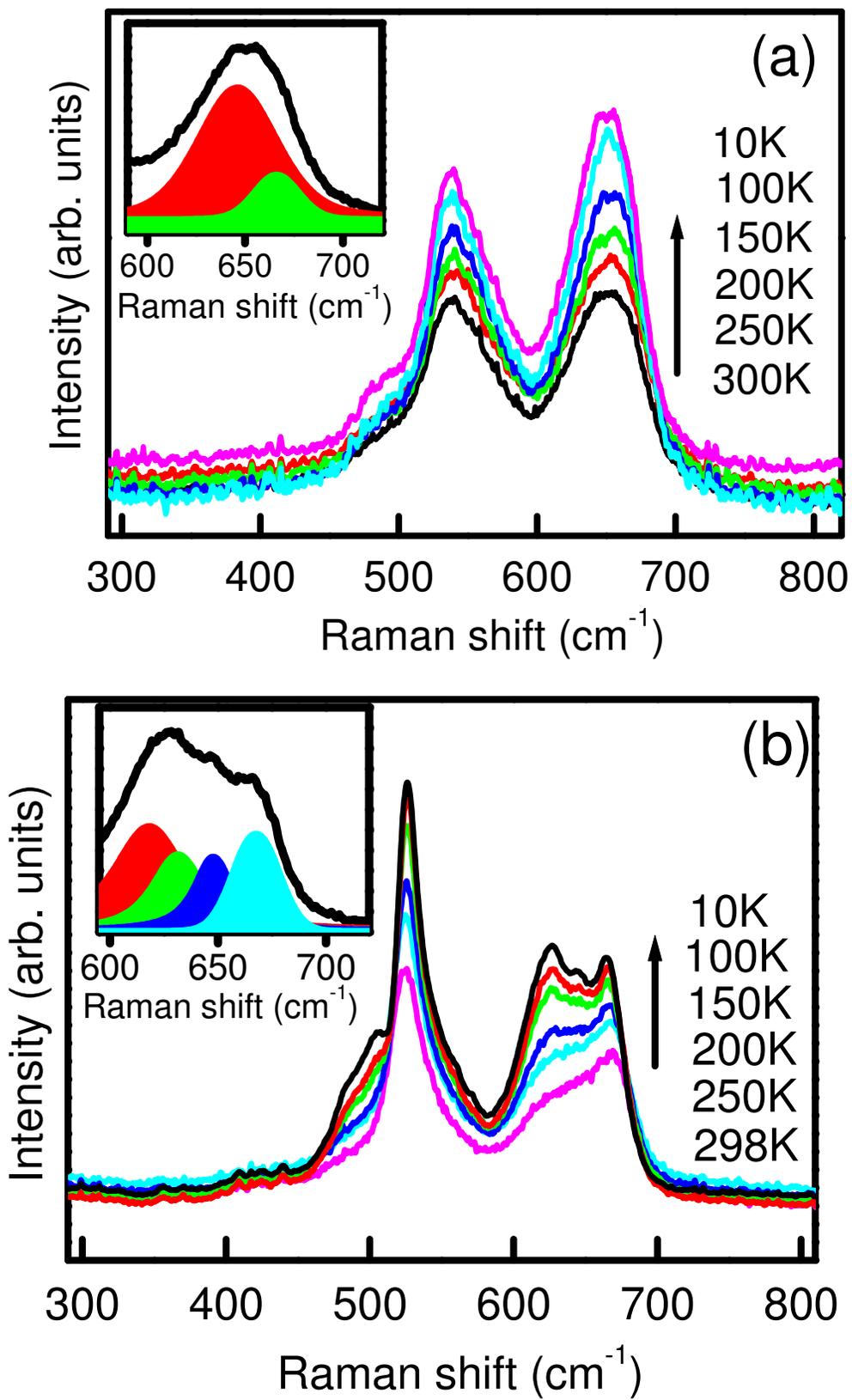

Figure 3

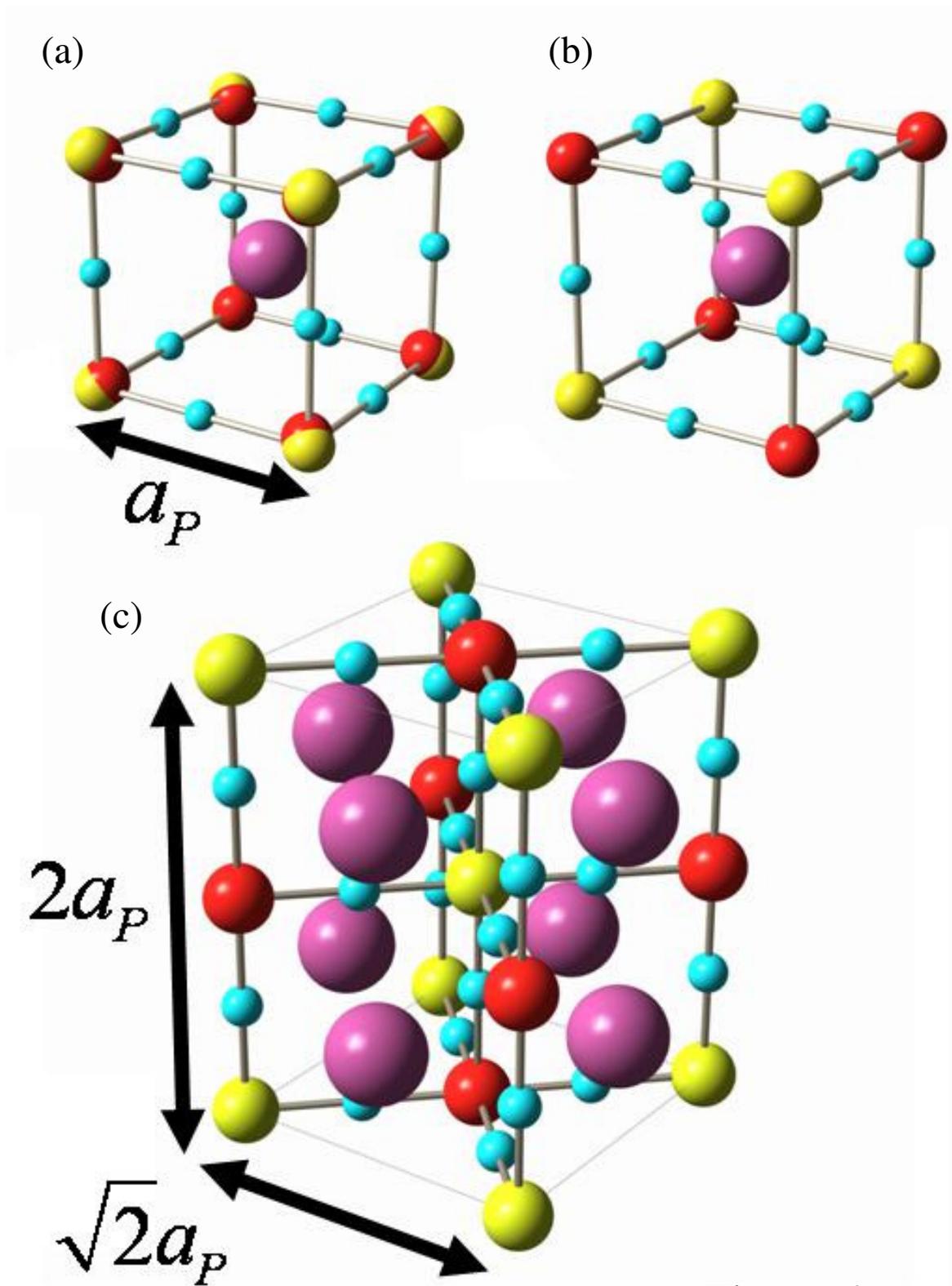

Figure 4

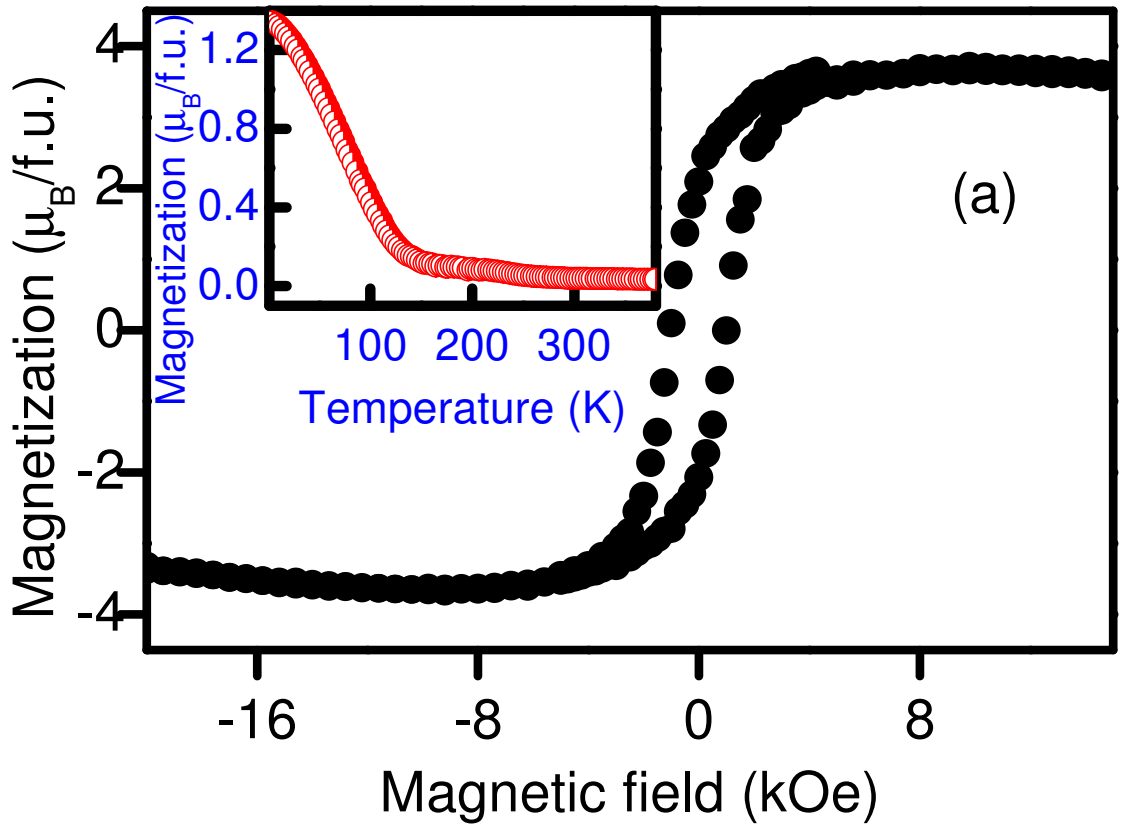
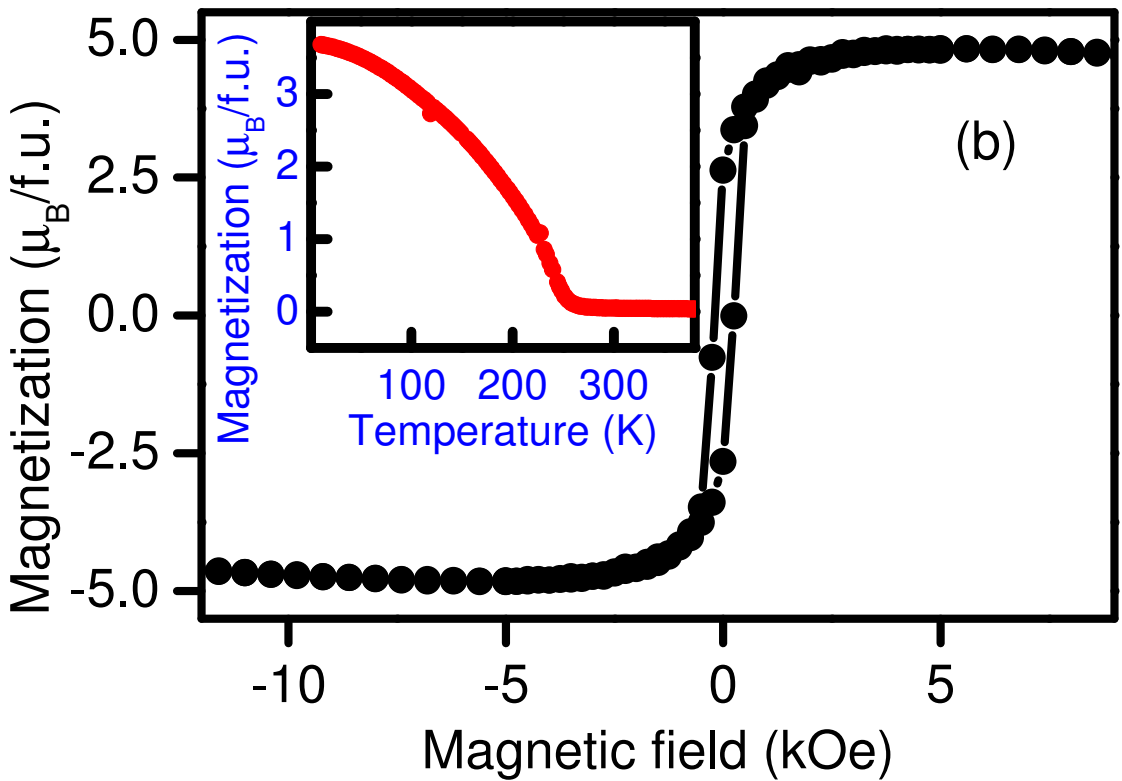

Figure 5